\documentclass[a4paper, amsfonts, amssymb, amsmath, reprint, showkeys, footinbib, ,twoside,superscriptaddress,floatfix,longbibliography]{revtex4-2}
\usepackage[USenglish]{babel}

\usepackage{graphicx} 
\usepackage{amsmath}
\usepackage{hyperref}
\usepackage{xr}
\usepackage{comment}

\usepackage{booktabs}
\usepackage{hyperref}
\usepackage{url}
\usepackage{adjustbox}
\usepackage{array}
\usepackage{xcolor}

\newcommand{\figLabelCapt}[1]{\textbf{\MakeLowercase{{#1}}}}
\newcommand{\refSub}[2]{\hyperref[#2]{\ref{#2}\figLabelCapt{#1}}}

\newcommand{\br}[1]{\mathbf{r}}
\newcommand{\bk}[1]{\mathbf{k}}

\begin{document}

\author{Xintian Wang}
\affiliation{Department of Materials Science and Engineering, National University of Singapore, Singapore 117575, Singapore}

\author{Junmin Chen}
\email[]{jmchen@nus.edu.sg}
\affiliation{Department of Materials Science and Engineering, National University of Singapore, Singapore 117575, Singapore}

\author{Zhuoying Zhu}
\affiliation{State Key Laboratory of Precision and Intelligent Chemistry, University of Science and Technology of China, Hefei, Anhui, 230026, China}

\author{Peichen Zhong}
\email[]{zhongpc@nus.edu.sg}
\affiliation{Department of Materials Science and Engineering, National University of Singapore, Singapore 117575, Singapore}



\title{Differentiable hybrid force fields support scalable autonomous electrolyte discovery}

\date{\today}

\begin{abstract}
Autonomous electrolyte discovery demands a computational engine that satisfies a critical trilemma: it must be fast enough for high-throughput screening, accurate enough for quantitative property prediction, and calibratable enough for online refinement. Classical empirical force fields (FFs) are fast but rely on error cancellation, while standard machine learning interatomic potentials (MLIPs) are computationally expensive. In this Perspective, we highlight that differentiable hybrid FFs resolve this trilemma by fusing physically motivated functional forms with neural-network short-range corrections. Grounded in Energy Decomposition Analysis (EDA), state-of-the-art models such as PhyNEO-Electrolyte and ByteFF-Pol achieve zero-shot generalization to bulk phases, delivering throughputs on the order of tens of ns/day (up to $\sim$50 ns/day, depending on model complexity) for 10,000-atom systems. Crucially, their physical skeletons provide a well-conditioned parameter space for differentiable molecular dynamics (dMD). This enables a dual-calibration paradigm: bottom-up \textit{ab initio} parameterization combined with top-down fine-tuning from macroscopic experimental observables. We propose that this architecture meets the requirements of a ``ChemRobot-ready'' digital twin by integrating physics-grounded simulation with experimentally calibratable refinement, thereby enabling closed-loop autonomous electrolyte discovery.

\end{abstract}

\maketitle


\section{Introduction}

\begin{figure*}[htb]
  \centering
  \includegraphics[width=1.0\linewidth]{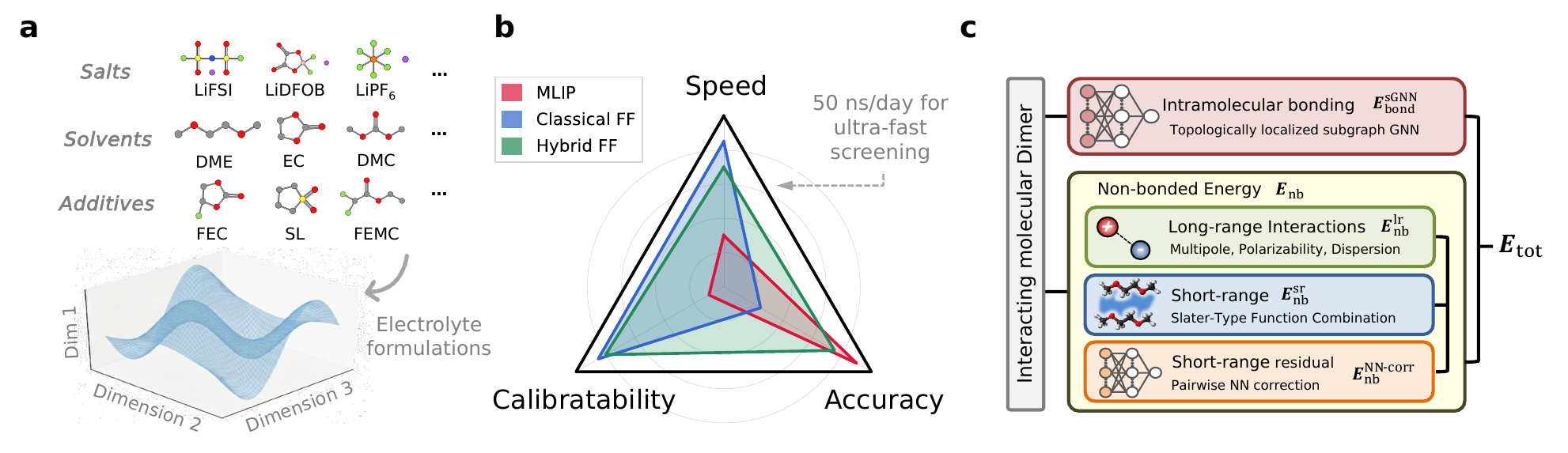}
  \caption{(a) Representative electrolyte design space: salts, solvents, and additives spanning the combinatorial formulation landscape. (b) Schematic radar plot comparing classical FFs, standard MLIPs, and differentiable hybrid FFs across the speed–accuracy–calibratability trilemma. Throughputs approaching $\sim$50 ns/day mark the ultra-fast screening regime. (c) Hierarchical energy decomposition of the hybrid FF. The total energy separates into intramolecular bonding ($E_{\mathrm{bond}}^{\mathrm{sGNN}}$, subgraph GNN) and non-bonded interactions ($E_{\mathrm{nb}}$), with the latter further partitioned into long-range Ewald summation ($E_{\mathrm{nb}}^{\mathrm{lr}}$; multipole electrostatics, polarization, dispersion), short-range Slater-type functions ($E_{\mathrm{nb}}^{\mathrm{sr}}$), and a pairwise neural-network correction ($E_{\mathrm{nb}}^{\mathrm{NN\text{-}corr}}$).
  }
  \label{fig:trilemma}
\end{figure*}

The rational design of liquid electrolytes for next-generation batteries confronts a combinatorial explosion: the space of solvents, salts, and additives is too vast to navigate by trial-and-error experiments alone \cite{hannah2025searching, kim_high-entropy_2023} (Figure~\ref{fig:trilemma}a). 
Molecular dynamics (MD) simulations serve as a natural high-throughput surrogate for these physical experiments~\cite{Yao2022_ChemRev, Xu2004_ChemRev, Meng2022_Science}.
However, the reliability of MD largely depends on the quality of the underlying force field (FF), which must satisfy a trilemma of speed, accuracy, and calibratability, as illustrated in Figure~\ref{fig:trilemma}b. 
\emph{Speed} is essential, as screening thousands of formulations demands tens of nanoseconds per day on accessible hardware. 
\emph{Accuracy} is crucial because macroscopic transport coefficients are nonlinear amplifiers of potential energy surface (PES) gradients; even a minor bias in the repulsion slope or polarization response can produce order-of-magnitude errors in bulk dynamics~\cite{Bedrov2019_ChemRev}. 
Finally, \emph{calibratability} is necessary because no \textit{ab initio} model is perfect. 
Deploying such a digital twin requires that the model remain aligned with experimental measurements without sacrificing physical stability.

Nonetheless, no existing paradigm satisfies all three requirements simultaneously. Classical empirical FFs such as OPLS-AA~\cite{Jorgensen1996_JACS} and GAFF~\cite{Wang2004_JCC} are fast and stable, but rely on experimental fitting and error cancellation~\cite{Chen2025_PCCP_review}. 
Machine learning interatomic potentials (MLIPs) achieve near-quantum accuracy but are typically more than $20\times$ slower at comparable system sizes.
Notably, the standard short-range MLIPs lack explicit long-range electrostatics and polarization \cite{Anstine2023_JPCA_longrange, Yue2021_JCP_shortrange, Kim2025_LES_Universal_JCTC}, and are numerically ill-conditioned for gradient-based calibration to macroscopic observables~\cite{Chen2025_PCCP_review, Fu2023_arXiv}. 
We envision that differentiable hybrid FFs \cite{Chen2024_PhyNEO, Chen2026_PhyNEO_Electrolyte, Zheng2025_ByteFFPol}, exemplified by PhyNEO-Electrolyte \cite{Chen2026_PhyNEO_Electrolyte} and ByteFF-Pol \cite{Zheng2025_ByteFFPol}, offer a principled resolution: by grounding the PES in EDA-decomposed physical components~\cite{Schmidt2015_AccChemRes}, one can (i) achieve zero-shot transferable accuracy (i.e., predictability of bulk-phase observables for unseen molecules, formulations, and concentration regimes composed of atom types and local chemical environments represented in the quantum-chemistry training data without experimental data alignment), (ii) retain the throughput of a semi-analytical model, and (iii) expose a well-conditioned parameter space for both bottom-up and top-down calibration. Benchmarks on multicomponent electrolyte formulations indicate that hybrid FFs outperform other current model classes on the joint speed--accuracy--calibratability profile required for autonomous screening~\cite{Chen2026_PhyNEO_Electrolyte, Zheng2025_ByteFFPol}, which we elaborate in Section II.

Concretely, the hybrid architecture (here exemplified by PhyNEO-Electrolyte) decomposes the total energy into nonbonded and bonded contributions
\begin{equation}
\begin{aligned}
        & E_{\mathrm{total}} =
E_{\mathrm{nb}} +
E_{\mathrm{bond}}^{\mathrm{sGNN}}\\
        & E_{\mathrm{nb}} =
E_{\mathrm{nb}}^{\mathrm{lr}} +
E_{\mathrm{nb}}^{\mathrm{sr}} +
E_{\mathrm{nb}}^{\mathrm{NN\text{-}corr}}
\end{aligned}
\label{ff_1}
\end{equation}
where $E_{\mathrm{nb}}^{\mathrm{lr}}$ captures long-range electrostatics, polarization, and dispersion via Ewald summation; $E_{\mathrm{nb}}^{\mathrm{sr}}$ models short-range interactions through Slater-type functions; $E_{\mathrm{nb}}^{\mathrm{NN\text{-}corr}}$ is a pairwise neural-network correction for short-range residuals; and $E_{\mathrm{bond}}^{\mathrm{sGNN}}$ handles intramolecular bonding via a subgraph GNN~\cite{Wang2021_JPCL_sGNN} (see Figure~\ref{fig:trilemma}c). 
In ByteFF-Pol, an analogous decomposition replaces the Slater functions with a modified Buckingham potential and adds an explicit charge-transfer term, with all non-bonded parameters predicted by a graph neural network trained against ALMO-EDA labels~\cite{Khaliullin2007_JPCA_ALMOEDA, Zheng2025_ByteFFPol}. 
This physical skeleton encodes the correct long-range asymptotics and short-range repulsive wall. 
Interactions therefore exhibit the physically expected electrostatic and dispersion decay at long range, while the repulsive wall prevents atoms from collapsing into unphysical close-contact configurations at short distances. The neural components then capture residual anisotropy and charge penetration, such as lone-pair interactions, which are difficult to represent analytically. Thus, hybrid FFs should be viewed not simply as reparameterized empirical FFs, but as a distinct model class that combines physically constrained interaction terms with neural residual corrections.

More importantly, the hybrid architecture is constructed from differentiable functional forms \cite{jax2018github}, enabling gradient-based optimization of the FF parameters~$\theta$ with respect to experimental data.
This opens the door to \textit{top-down calibration}: the FF drives an MD simulation, predicted observables are compared against experimental measurements, and the discrepancy is used to update~$\theta$ by minimizing a loss of the form
\begin{equation}
  \mathcal{L}(\theta)
    = \frac{1}{K}\sum_{k=1}^{K}
      \bigl[\langle O_k(U_\theta)\rangle
            - \tilde{O}_k\bigr]^2,
\end{equation}
where $\langle O_k(U_\theta)\rangle$ is the ensemble average of observable~$O_k$ computed from the potential~$U_\theta$, and $\tilde{O}_k$ is the corresponding experimental measurement. While many MLIPs are also formally differentiable, their high-dimensional parameter spaces can make gradient-based calibration more challenging in practice. In contrast, the relatively low-dimensional and physically structured parameterization of hybrid FFs makes such optimization more tractable~\cite{Han2025_NatCommun_IR}.

The remainder of this Perspective elaborates on each vertex of the trilemma and argues that their simultaneous resolution constitutes the correct design criterion for FFs that are ``ChemRobot-ready'' and can serve as the computational engine of an autonomous electrolyte discovery laboratory.

\section{What makes a potential ChemRobot-ready?}

Figure~\ref{fig:trilemma}b presents the three mutually enabling vertices of this trilemma. Computational speed is a strict prerequisite for calibration, as differentiable molecular dynamics (dMD) requires trajectories long enough to converge observable gradients---a condition that is practical at tens of ns/day (up to $\sim$50 ns/day depending on model complexity) but prohibitive at only a few ns/day. 
The EDA formalism renders this calibration highly tractable by exposing a low-dimensional, well-conditioned parameter space, in contrast to the millions of opaque weights in deep neural networks. This robust calibration in turn closes the accuracy gap left by purely bottom-up construction, allowing experimental data to directly inform the next screening cycle. Consequently, any FF lacking even one of these three vertices will inevitably introduce a computational or predictive bottleneck. We elaborate on each below.

\subsection{Speed: Efficiency without physical compromise}

Converged transport properties require trajectories spanning tens to hundreds of nanoseconds in simulation cells containing thousands of atoms, and a virtual screening campaign must repeat this across many formulations. Against this benchmark, current MLIPs fall significantly short. On a 10,000-atom electrolyte system on a single NVIDIA RTX 5090, PhyNEO-Electrolyte achieves approximately tens of ns/day (up to $\sim$50 ns/day, depending on model complexity)~\cite{Chen2026_PhyNEO_Electrolyte}, more than $20\times$ faster than state-of-the-art message-passing MLIPs such as MACE-OFF23~\cite{Kovacs2025_MACEOFF_JACS} and BAMBOO~\cite{Gong2025_BAMBOO_NMI} at comparable accuracy. ByteFF-Pol reports a similar throughput of $\sim$50~ns/day on an NVIDIA L20 GPU~\cite{Zheng2025_ByteFFPol}.

The efficiency advantage is architectural. Deep message-passing networks require many-body feature aggregation at every atom, which is expensive and difficult to parallelize for large periodic cells. The hybrid architecture instead separates interactions by range: long-range electrostatics and polarization are handled by Ewald summation~\cite{Essmann1995_JCP} and a dipole self-consistency loop~\cite{Huang2017_JCP_MPID}, while the neural component is restricted to a short-range correction and bonding interaction. This leads to a substantially shallower computational graph than a full message-passing MLIP~\cite{musaelian_learning_2023}, while retaining richer physics than classical FFs. 

Beyond efficiency, the physical skeleton provides the correct long-range asymptotics and repulsive wall that keep simulations stable over hundreds of nanoseconds. This is a prerequisite for converging transport properties, whereas many standard MLIPs exhibit instability and unphysical drift in this regime~\cite{Fu2023_arXiv, Chen2025_PCCP_review}.

\begin{figure*}[htb]
    \centering
    \includegraphics[width=1\linewidth]{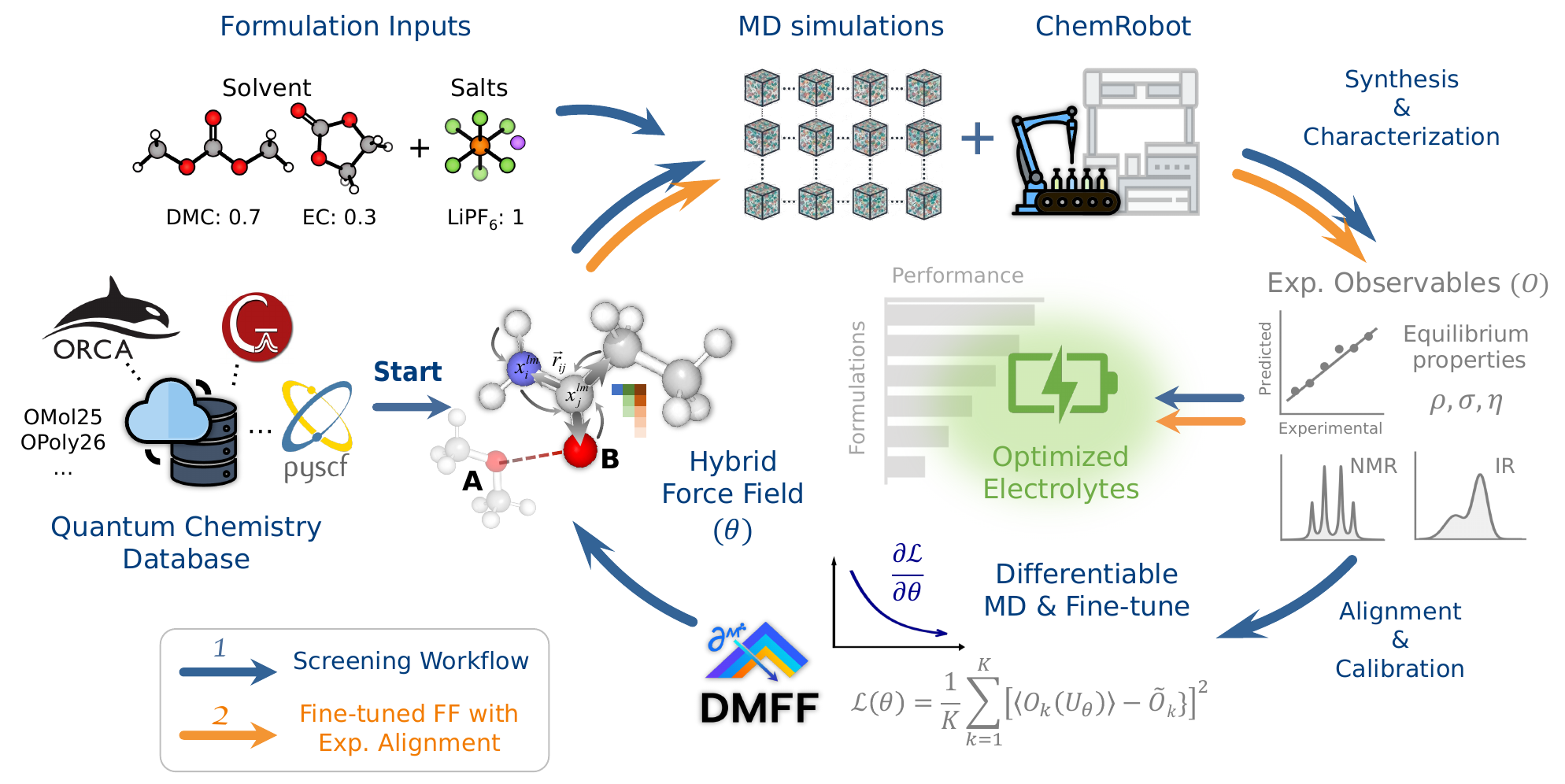}
    \caption{Computational-experimental workflow for autonomous electrolyte discovery. The workflow begins with the hybrid FF ($\theta$), parameterized bottom-up from quantum chemistry databases. (1, blue arrows) In the screening workflow, formulation inputs (e.g., DMC:EC:LiPF$_6$ at specified ratios) are simulated via high-throughput MD; candidates ranked by predicted performance are passed to the ChemRobot platform for synthesis and electrochemical characterization. The resulting experimental observables (equilibrium properties $\rho$, $\sigma$, $\eta$, and spectroscopic signatures such as NMR and IR) then drive top-down calibration via differentiable MD (e.g., DMFF), updating the FF parameters $\theta$ by minimizing the loss $\mathcal{L}(\theta)$. (2, orange arrows) Subsequent screening cycles use the fine-tuned FF, now quantitatively aligned with experimental measurements.
    Molecular structures are visualized using \texttt{xyzrender} \cite{goodfellow_graph-based_2026}.
    }
    \label{fig:work_flow}
\end{figure*}

\subsection{Accuracy: From dimer physics to bulk predictability}

We conceptualize accuracy as the transferable bulk predictability rather than agreement with in-distribution bulk benchmarks alone. In hybrid FFs, this transferability comes from EDA-grounded dimer fitting: EDA resolves the dominant non-bonded interactions into physical components with the correct asymptotic behavior, allowing much of the short-range energy to be represented by compact pairwise forms such as Slater-type functions \cite{Chen2026_PhyNEO_Electrolyte} or modified Buckingham functions \cite{Zheng2025_ByteFFPol}, and leaving only a smaller residual for the neural correction.

\textbf{Dimer interactions as transferable training targets.}~
A ChemRobot-ready potential must maintain predictive accuracy across diverse condensed-phase environments. 
On bulk benchmarks within the training distribution, current MLIPs can match or slightly exceed hybrid models. This comparison is, however, misleading for two reasons. 
First, bulk-fitted MLIPs require large condensed-phase \textit{ab initio} datasets, typically from expensive AIMD trajectories, whereas the hybrid framework achieves comparable bulk performance from dimer-level quantum chemistry without requiring bulk data, reducing training cost by an order of magnitude~\cite{Chen2026_PhyNEO_Electrolyte, Zheng2025_ByteFFPol}. Second, bulk accuracy from bulk fitting does not transfer: accuracy degrades unpredictably when the MLIP encounters a new solvent or concentration. 

This problem is pervasive across the MLIP landscape, spanning general-purpose, polarizable, and electrolyte-specific models alike. MACE-OFF23 overestimates the density of liquid water by approximately 20\% at its default 5~\AA{} cutoff (a consequence of missing long-range electrostatic contributions beyond the receptive field) and, even with an extended 6~\AA{} cutoff, the error remains 2--5\% across the full temperature range~\cite{Kovacs2025_MACEOFF_JACS}. The recently released state-of-the-art MACE-POLAR-1~\cite{Batatia2026_MACEPOLAR1}, which adds explicit electrostatic induction to the MACE architecture, still overestimates water density by $\sim$10\%, whereas the physically motivated MB-pol achieves errors below 1\%~\cite{Zhu2023_MBpol2023}. 
Among electrolyte-focused MLIPs, BAMBOO requires an empirical ``density alignment'' step using experimental data to correct systematic density biases~\cite{Gong2025_BAMBOO_NMI}; without this post-hoc correction, its zero-shot bulk predictions are substantially worse. 
More broadly, independent benchmarks have shown that current transferable NNPs (including ANI-2x and MACE-OFF23 variants) can overestimate liquid densities, yield unphysical isothermal compressibilities, and produce dramatically slowed self-diffusion, sometimes rendering aqueous simulations entirely unusable~\cite{Beckmann2025_JCIM_NNP_condensed}. 
These failures are not simply a matter of insufficient training data but reflect fundamental architectural limitations. \citet{niblett_transferability_2025} demonstrated that MLIP stability for molecular liquids is conserved only for small changes in molecular shape but not for changes in functional chemistry, necessitating system-specific retraining. \citet{Yue2021_JCP_shortrange} further showed that short-range MLIPs systematically fail to reproduce bulk dielectric properties and vapor--liquid equilibria when electrostatic screening lengths exceed the cutoff~\cite{Anstine2023_JPCA_longrange}. These issues underscore the fundamental difficulty of extrapolating short-range, single-phase fitted potentials to long-range-dominated condensed-phase observables~\cite{Kim2025_LES_Universal_JCTC}.

In contrast, the hybrid framework delivers quantitative bulk predictions without any experimental fitting. PhyNEO-Electrolyte achieves a density error of 0.73\% across multicomponent Li- and Na-ion electrolyte formulations, outperforming the empirically fitted OPLS (1.11\%) and reproducing Li$^+$ diffusion coefficients within 10--20\% of experiment across temperatures and compositions~\cite{Chen2026_PhyNEO_Electrolyte}.
ByteFF-Pol predicts densities with a mean absolute percentage error of 3.0\% and evaporation enthalpies within 11.2\%, and achieves a Pearson correlation of 0.95 for conductivity across nearly 5,000 electrolyte systems~\cite{Zheng2025_ByteFFPol}. This bulk predictivity is rooted in the correct physical decomposition of intermolecular forces, which generalizes across chemical space without empirical post-hoc corrections.
In this sense, EDA is not merely a labeling strategy; it reduces the complexity of the target and is the reason dimer-level fitting can transfer to bulk environments.

\textbf{Energy decomposition as the data strategy for transferability.}~
The physical decomposition in Equations~\eqref{ff_1} directly dictates the training data requirements. Since each non-bonded component is assigned a functional form with the correct asymptotic behavior, the neural network needs only to learn residual short-range anisotropy from dimer-level quantum chemistry~\cite{VanVleet2018_JCTC_anisotropy}. 
Monomer properties (atomic multipoles, polarizabilities, dispersion coefficients) are obtained from ISA/ISA-pol~\cite{Misquitta2018_ISApol, Misquitta2014_JCTC_ISA} and TD-DFT calculations~\cite{McDaniel2012_JPC_zeolite}. Approximately 500,000 dimer data points suffice for broad coverage of chemical space~\cite{Chen2026_PhyNEO_Electrolyte}, far fewer than general many-body MLIPs require.

\textbf{Architectural constraints that preserve transferability.}~
The hybrid architecture achieves physical rigor through a clear separation of bonding and non-bonding interactions, combined with range separation. For intramolecular degrees of freedom, strictly localized sub-graph neural networks (sGNN)~\cite{Wang2021_JPCL_sGNN} or GNN-predicted bonded parameters~\cite{Zheng2025_ByteFF_DataDriven} are trained on single-molecule datasets, cleanly separated from the non-bonding interactions. For non-bonding interactions, the physical base model handles the dominant contributions, while the neural correction refines the short-range residual. 
This restriction of the neural component to a bounded short-range correction is not merely a design preference, but the mechanism that ensures the repulsive wall remains intact, preventing the unphysical close-contact configurations that destabilize long-time MD trajectories in standard MLIPs. Related hybrid architectures, including ARROW-NN~\cite{Illarionov2023_JACS_ARROW, Kamath2024_JPCA_ARROW}, FeNNol~\cite{Ple2024_JCP_FeNNol}, and the range-separated water models~\cite{Yang2022_JCP_water, Gao2026_arXiv_water}, share this design philosophy, confirming the generality of the approach.

\subsection{Calibratability: Differentiable MD and the dual feedback loop}

Speed and accuracy alone are insufficient for a ``ChemRobot-ready'' potential. A digital twin that cannot update itself as experimental data arrive will drift out of calibration when applied to novel formulations. This is the calibration requirement, where the differentiable nature of the hybrid architecture becomes decisive.

We distinguish two complementary calibration directions. \emph{Bottom-up calibration} proceeds from quantum mechanics to the FF: new molecules are characterized by EDA calculations on a small number of dimers, and the decomposed energies extend the model's coverage of chemical space. Because the physical skeleton already encodes the correct functional form, only the residual neural correction needs updating, requiring far fewer data points than retraining a standard MLIP~\cite{Chen2026_PhyNEO_Electrolyte}. 
\emph{Top-down calibration} proceeds in the opposite direction: macroscopic observables from the robotic platform (density and spectroscopic data) are used to compute gradients with respect to FF parameters via dMD, and the parameters are updated to reduce the discrepancy with experiment. Bottom-up calibration maintains physical grounding as new chemistries are encountered; top-down calibration corrects for systematic biases from DFT approximations and missing physical effects. 
Together, they form a dual feedback loop that continuously improves the FF as the ChemRobot accumulates experimental data.

The development of differentiable MD frameworks, including DMFF~\cite{Wang2023_JCTC_DMFF}, JAX-MD~\cite{Schoenholz2020_NeurIPS_JAXMD}, TorchMD~\cite{Doerr2021_JCTC_TorchMD}, DIMOS~\cite{christiansen2025_DIMOS}, and chemtrain~\cite{Fuchs2024_arXiv_chemtrain}, has made top-down calibration practically feasible, while the validation status of dMD varies across observable classes. For equilibrium thermodynamic and structural properties such as density, heat of vaporization, and the radial distribution function, differentiable reweighting estimators~\cite{Wang2023_JCTC_DMFF, Thaler2021_NatCommun} provide stable gradient signals and have been applied to electrolyte-relevant liquids~\cite{Wang2023_JCTC_DMFF}. Recent work has extended top-down calibration to dynamical observables:
\citet{Han2025_NatCommun_IR} demonstrated that infrared spectra can be differentiated along MD trajectories using adjoint methods and gradient truncation, enabling FF refinement from spectroscopic data. Electrolyte-specific validation of this route remains an open question. Transport coefficients represent a more challenging tier of dynamical targets because they require long-time correlation functions, where gradient explosion, vanishing, and noise accumulation over long trajectories remain unresolved. Currently developed stabilization strategies include gradient truncation at a decorrelation timescale, differentiation of the dynamical component of the gradient to reduce statistical noise~\cite{Han2025_NatCommun_IR}, and reversible-simulation schemes with effectively constant memory cost over long trajectories~\cite{Greener2025_PNAS_reversible}. Beyond liquid-state observables, phase diagrams~\cite{Jin2025_arXiv_phasediagram} and phase transition temperatures~\cite{Rocken2025_arXiv_DiffTTC} have also been used as differentiable calibration targets. These developments demonstrate that the dMD calibration loop is operational for equilibrium observables, with spectroscopic and dynamical extensions at an earlier stage of electrolyte-specific validation.

The hybrid architecture is also uniquely well-conditioned for dMD compared to standard MLIPs. A typical hybrid model exposes $\mathcal{O}(10^2)$ physically meaningful parameters (Slater exponents, damping coefficients, dispersion coefficients), whereas a message-passing MLIP contains $\mathcal{O}(10^6)$ opaque weights, whose high-dimensional parameter space leads to chaotic gradient landscapes when back-propagating through long MD trajectories~\cite{Metz2021_arXiv_gradients}. The physical repulsive wall further stabilizes gradient computation by preventing unphysical close-contact configurations during parameter perturbation. 
A rough estimate underscores the speed--calibration coupling: if each top-down update requires $\sim$10~ns of trajectory to converge the gradient of a transport property, and $\sim$100 updates are needed for convergence, the total cost is $\sim$1~$\mu$s of MD. This is where the throughput advantage of the hybrid architecture becomes decisive: at 50~ns/day this takes $\sim$20~days; at 2.5~ns/day it would take over a year.

\section{Toward ChemRobot Integration}

Resolving the trilemma can enable electrolyte-formulation screening at an unmatched scale by integrating the differentiable hybrid FF into an autonomous laboratory---``ChemRobot'' workflow (Figure~\ref{fig:work_flow}).
Bulk equilibrium properties are a meaningful and appropriate entry point for autonomous electrolyte screening, yet may be insufficient for a full electrolyte-design problem.
Autonomous experimental platforms such as Clio~\cite{Dave2022_NatCommun_robot} and DiffMix~\cite{Zhu2024_NatCommun_DiffMix} have already demonstrated closed-loop electrolyte optimization; however, their computational surrogates operate at the property-correlation level (e.g., machine-learning surrogates mapping composition to conductivity), lacking the atomistic resolution needed to diagnose \emph{why} a formulation fails or to extrapolate beyond the training distribution. The hybrid FF fills this gap precisely: it provides an atomistic-resolution digital twin whose parameters can be refined from the same experimental observables the robot measures, enabling mechanism-informed exploration of chemical space.

The envisioned ChemRobot workflow operates in two stages following bottom-up hybrid FF development. 
In the \emph{screening stage}, new formulations are assigned parameters through automated GNN-based parameterization~\cite{Zheng2025_ByteFFPol, Zheng2025_ByteFF_DataDriven} or direct physical transfer from the existing molecular library~\cite{Chen2026_PhyNEO_Electrolyte}, and bulk MD predicts target properties (conductivity, viscosity, density) within hours. Redox potentials can be accurately evaluated using foundation potentials and quantum chemical methods~\cite{Chen2026_PrecChem_redox}. Promising candidates pass to the robotic platform for synthesis and measurement.
In the \emph{refinement stage}, experimental results trigger top-down dMD calibration, including density via differentiable thermodynamic reweighting~\cite{Wang2023_JCTC_DMFF, Thaler2021_NatCommun}, and spectroscopic data via adjoint-based trajectory differentiation~\cite{Han2025_NatCommun_IR}; transport properties such as viscosity and conductivity are longer-term calibration targets as gradient-conditioning methods for long-time dynamics mature. 

Essentially, spectroscopy provides the critical bridge between microscopic FF parameters and macroscopic measurable properties. 
Any macroscopic observable corresponds, at the atomistic level, to either a thermodynamic ensemble average or a time-correlation function~\cite{McQuarrie1976_StatMech}. For instance, the infrared spectrum (IR) is the Fourier transform of the dipole autocorrelation function, NMR relaxation rates encode rotational dynamics, and Raman spectra probe polarizability fluctuations~\cite{Zhang2020_EANN_spectra, Cheng2024_JCTC_equivariant}. These quantities are experimentally accessible on automated platforms and computable from MD trajectories~\cite{Han2025_NatCommun_IR}, yet converging the relevant time-correlation functions typically requires tens of nanoseconds of NVE dynamics. A FF that simultaneously reproduces spectroscopic signatures and thermodynamic observables will provide the most stringent ``genome-to-function'' validation of the underlying PES~\cite{Han2025_NatCommun_IR, zhong_machine_2025}. The throughput of the hybrid FF makes such closed-loop workflows practically viable, with calibration costs amortized across successive candidates as the FF matures.

Recent research practice suggests that ChemRobot integration is most effective when formulated as a hierarchical decision architecture that couples atomistic simulation with robotic experimentation \cite{zhang_revolutionizing_2025}. Within this architecture, the differentiable hybrid FF functions as the mechanistic core: it maps formulation inputs to transport coefficients and spectroscopic signatures that are directly comparable with robotic measurements, thereby providing a shared representational language between simulation and experiment.
Lessons from recent closed-loop studies reinforce this point: theory is most valuable when it goes beyond front-end candidate filtering and actively guides iteration throughout the discovery cycle \cite{shen_unlocking_2026, sun_moses_2026, song_multiagent-driven_2025}---identifying unconverged regions, triggering follow-up calculations, and supplying physically grounded descriptors for experimental acquisition functions.
However, coordinating these heterogeneous information streams demands context-dependent judgment that cannot be captured by a fixed protocol. This is precisely where an agentic orchestration layer becomes essential: it must decide, at each iteration, whether the current uncertainty is best resolved by launching another simulation or by  allocating a robotic slot, and it must reconcile feedback of fundamentally different fidelity and cost.
In this view, a ChemRobot-ready workflow is best understood as a model-centric ecosystem in which the hybrid FF is continuously calibrated by observables at equilibrium, the agentic layer manages the simulation--experiment interface, and the resulting experimentally grounded surrogate gradually evolves from a task-specific predictor into a reusable digital twin for autonomous discovery.

Nonetheless, open challenges remain at multiple scales. At the FF level, 
anisotropic interactions such as halogen bonds and $\pi$-stacking require conformation-dependent atomic multipoles for accurate description, a challenge that can be addressed through equivariant machine learning without sacrificing physical interpretability~\cite{Cheng2024_JCTC_equivariant}. 
Many-body polarization in concentrated aqueous electrolytes and in interfacial double layers also remains an active development area~\cite{Gao2026_arXiv_water}. At the measurement level, interfacial layering, concentration gradients, and evolving electrode--electrolyte configurations can distort the ideal response computed from homogeneous MD~\cite{Ye2026_Interfacial}. This makes top-down calibration from cell-level spectroscopy more difficult, as matching the experimental geometry is a prerequisite for quantitatively interpreting the differentiated observable. At the workflow level, bridging FF predictions to cell-level performance requires multiscale coupling with continuum transport models. For example, MD-predicted properties such as ion diffusion coefficients, transference numbers, and viscosities can serve as inputs to porous-electrode models (e.g., Doyle–Fuller–Newman-type battery models) or to Nernst–Planck transport equations that describe ion transport at the device scale. The community still lacks standardized benchmarks for comparing FF accuracy on electrolyte-relevant properties across diverse chemical spaces~\cite{Chen2025_PCCP_review}. These extensions are technically demanding but conceptually natural within the hybrid framework.

\section{Discussion and Outlook}

In this Perspective, we propose that the key design target for next-generation electrolyte FFs is the simultaneous satisfaction of speed, accuracy, and calibratability. 
This trilemma is especially important for autonomous electrolyte discovery, where the computational model must efficiently screen large chemical spaces, quantitatively predict bulk behavior, and remain updatable as new experimental measurements become available. 
Differentiable hybrid FFs currently represent the most promising route toward a deployable atomistic engine. Their advantage is architectural. By combining a physically grounded long-range framework with learned short-range corrections, hybrid FFs retain the stability, interpretability, and efficiency of analytical models while recovering the accuracy and transferability typically associated with machine-learned potentials. 
Models such as PhyNEO-Electrolyte~\cite{Chen2026_PhyNEO_Electrolyte} and ByteFF-Pol~\cite{Zheng2025_ByteFFPol} illustrate that this combination can already deliver zero-shot transfer from dimer-level training to bulk liquid properties at practically relevant throughput. Equally important, their structured parameterization exposes a well-conditioned space for dMD, allowing top-down refinement against thermodynamic, dynamical, and spectroscopic observables without losing physical meaning~\cite{Wang2023_JCTC_DMFF, Thaler2021_NatCommun}. 
This stands in contrast to classical empirical FFs, whose parameters can also be fitted to experiment but whose reliance on error cancellation renders such fitting non-transferable across chemistries and ill-suited to the multi-objective refinement demanded by autonomous workflows.

Beyond static property prediction, the high throughput of hybrid FFs positions them as natural computational engines for autonomous electrolyte discovery when coupled with LLM-based agents. A promising closed-loop agentic workflow can be envisioned as follows: starting from MD agents that instantiate and orchestrate simulation workflows with minimal human intervention \cite{campbell_mdcrow_2026, ding_topolyagent_2026, guilbert_dynamate_2025, shi_mdagent2_2026}, progressing to active learning loops where surrogate models with calibrated uncertainty quantification guide simulation-in-the-loop validation to concentrate computational effort where information gain is greatest \cite{liularge2024,kristiadi2024sober}, advancing to optimization agents that navigate vast electrolyte composition spaces under specified computational objectives \cite{wangefficient2025,gan2025large}, and ultimately automating the discovery process by dynamically steering the objectives themselves -- from adjusting electrochemical boundary conditions and balancing multi-objective trade-offs to revising optimization formulations on the fly \cite{du2025accelerating}. Throughout the entire workflow, hybrid FFs provide fast, robust reward signals that close the loop, enabling agents to iteratively propose, simulate, and optimize electrolyte candidates. 
Crucially, a hybrid FF that continuously absorbs experimental feedback evolves from a static surrogate into a digital twin whose interactions are progressively corrected as new formulations are explored. As this in-silico pipeline matures, it will provide the computational backbone for the experiment-facing ChemRobot workflows envisioned in this Perspective.

Finally, several challenges still need to be addressed before this vision becomes routine. At the FF level, anisotropic interactions, interfacial chemistry, and many-body polarization remain incompletely described, particularly for concentrated electrolytes and electrochemical environments~\cite{VanVleet2018_JCTC_anisotropy, Ye2026_Interfacial, Gao2026_arXiv_water}. 
At the calibration level, the fundamental challenge extends beyond merely matching individual observables. It requires integrating diverse, often competing experimental feedback into FF updates that maintain physical interpretability and transferability across different compositions, concentrations, and operating conditions \cite{feng2025screening, Han2025_NatCommun_IR}.
At the workflow level, atomistic predictions must be connected more systematically to continuum transport and cell-scale performance models, and the field still lacks standardized benchmarks for evaluating FF quality across chemically diverse electrolyte systems.

Even with these open questions, the broader direction is clear. The most useful FFs for autonomous discovery will be the ones that best balance physical structure, predictive accuracy, computational throughput, and experimental updateability. We therefore expect differentiable hybrid FFs to serve as the computational foundation of future autonomous electrolyte-design platforms.

\section*{Acknowledgements} 
This work was supported in part by the AI2050 program at Schmidt Sciences (Grant G-25-69776).
J.C. acknowledges the support from the NUS-AISI Joint Research Initiative Fund. Z.Z. acknowledges the National Natural Science Foundation of China (Grant 52541002) and the University of Science and Technology of China (USTC) Startup Programs for funding. The authors thank Kuang Yu and Sang Cheol Kim for valuable discussions.

\bibliography{refs}

@article{Chen2024_PhyNEO,
  author    = {Chen, Junmin and Yu, Kuang},
  title     = {{PhyNEO}: A Neural-Network-Enhanced Physics-Driven Force Field Development Workflow for Bulk Organic Molecule and Polymer Simulations},
  journal   = {J. Chem. Theory Comput.},
  year      = {2024},
  volume    = {20},
  pages     = {253--265},
  doi       = {10.1021/acs.jctc.3c01045},
}

@article{Chen2026_PhyNEO_Electrolyte,
    title = {A {Hybrid} {Physics}-{Driven} {Neural} {Network} {Force} {Field} for {Liquid} {Electrolytes}},
    volume = {22},
    copyright = {https://doi.org/10.15223/policy-029},
    issn = {1549-9618, 1549-9626},
    url = {https://pubs.acs.org/doi/10.1021/acs.jctc.5c02100},
    doi = {10.1021/acs.jctc.5c02100},
    number = {6},
    urldate = {2026-03-29},
    journal = {Journal of Chemical Theory and Computation},
    author = {Chen, Junmin and Gao, Qian and Lin, Yange and Huang, Miaofei and Cheng, Zheng and Feng, Wei and Huang, Jianxing and Wang, Bo and Yu, Kuang},
    month = mar,
    year = {2026},
    pages = {3011--3022},
}

@article{Zheng2025_ByteFFPol,
	title = {Bridging quantum mechanics to liquid properties via a universal organic force field},
	copyright = {2026 The Author(s)},
	issn = {2041-1723},
	url = {https://www.nature.com/articles/s41467-026-73566-3},
	doi = {10.1038/s41467-026-73566-3},
	urldate = {2026-06-06},
	journal = {Nature Communications},
	publisher = {Nature Publishing Group},
	author = {Zheng, Tianze and Xu, Xingyuan and Wang, Zhi and Yang, Zhenze and Wang, Yuanheng and Han, Xu and Chen, Lei and Mu, Zhenliang and Zhang, Ziqing and Liu, Siyuan and Gong, Sheng and Yu, Kuang and Yan, Wen},
	month = may,
	year = {2026},
}

@article{Chen2025_PCCP_review,
  author    = {Chen, Junmin and Gao, Qian and Huang, Miaofei and Yu, Kuang},
  title     = {Application of Modern Artificial Intelligence Techniques in the Development of Organic Molecular Force Fields},
  journal   = {Phys. Chem. Chem. Phys.},
  year      = {2025},
  volume    = {27},
  pages     = {2294--2319},
  doi       = {10.1039/d4cp02989e},
}

@article{Illarionov2023_JACS_ARROW,
  author    = {Illarionov, Alexey and Sakipov, Serzhan and Pereyaslavets, Leonid and Kurnikov, Igor V. and Kamath, Ganesh and Butin, Oleg and Voronina, Ekaterina and Ivahnenko, Igor and Leontyev, Igor and Nawrocki, Grzegorz and Darkhovskiy, Mikhail and Olevanov, Mikhail and Cherniavskyi, Yuriy K. and Lock, Christopher and Greenslade, Simon and Sankaranarayanan, Subramanian K. R. S. and Kurnikova, Maria G. and Potoff, Jeffrey and Kornberg, Roger D. and Levitt, Michael and Fain, Boris},
  title     = {Accurate Representation of Intermolecular Interactions through Combining Force Fields and Neural Networks},
  journal   = {J. Am. Chem. Soc.},
  year      = {2023},
  volume    = {145},
  pages     = {23620--23629},
  doi       = {10.1021/jacs.3c07628},
}

@article{Kamath2024_JPCA_ARROW,
  author    = {Kamath, Ganesh and Illarionov, Alexey and Sakipov, Serzhan and Pereyaslavets, Leonid and Kurnikov, Igor V. and Butin, Oleg and Voronina, Ekaterina and Ivahnenko, Igor and Leontyev, Igor and Nawrocki, Grzegorz and Darkhovskiy, Mikhail and Olevanov, Mikhail and Cherniavskyi, Yuriy K. and Lock, Christopher and Greenslade, Simon and Chen, Yinglong and Kornberg, Roger D. and Levitt, Michael and Fain, Boris},
  title     = {Combining Force Fields and Neural Networks for an Accurate Representation of Bonded Interactions},
  journal   = {J. Phys. Chem. A},
  year      = {2024},
  volume    = {128},
  pages     = {807--812},
  doi       = {10.1021/acs.jpca.3c06614},
}

@article{Ple2024_JCP_FeNNol,
  author    = {Pl{\'e}, Thomas and Lagard{\`e}re, Louis and Piquemal, Jean-Philip},
  title     = {{FeNNol}: An Efficient and Flexible Library for Building Force-Field-Enhanced Neural Network Potentials},
  journal   = {J. Chem. Phys.},
  year      = {2024},
  volume    = {161},
  pages     = {042502},
  doi       = {10.1063/5.0216015},
  note      = {arXiv:2301.08734},
}

@article{Yang2022_JCP_water,
  author    = {Yang, Lan and Li, Jichen and Chen, Feiyang and Yu, Kuang},
  title     = {A Transferrable Range-Separated Force Field for Water: Combining the Power of Both Physically-Motivated Models and Machine Learning Techniques},
  journal   = {J. Chem. Phys.},
  year      = {2022},
  volume    = {157},
  pages     = {214108},
  doi       = {10.1063/5.0131544},
}

@article{Gao2026_arXiv_water,
  author    = {Gao, Qian and Chen, Junmin and Yu, Kuang},
  title     = {Refinement and Performance Benchmark for Range-Separated Water Force Field},
  journal   = {arXiv preprint},
  year      = {2026},
  eprint    = {2601.18416},
  archiveprefix = {arXiv},
}

@article{Jorgensen1996_JACS,
  author    = {Jorgensen, William L. and Maxwell, David S. and Tirado-Rives, Julian},
  title     = {Development and Testing of the {OPLS} All-Atom Force Field on Conformational Energetics and Properties of Organic Liquids},
  journal   = {J. Am. Chem. Soc.},
  year      = {1996},
  volume    = {118},
  pages     = {11225--11236},
  doi       = {10.1021/ja9621760},
}

@article{Wang2004_JCC,
  author    = {Wang, Junmei and Wolf, Romain M. and Caldwell, James W. and Kollman, Peter A. and Case, David A.},
  title     = {Development and Testing of a General Amber Force Field},
  journal   = {J. Comput. Chem.},
  year      = {2004},
  volume    = {25},
  pages     = {1157--1174},
  doi       = {10.1002/jcc.20035},
}

@article{Kovacs2025_MACEOFF_JACS,
  author    = {Kov{\'a}cs, D{\'a}vid P{\'e}ter and Moore, J. Harry and Browning, Nicholas J. and Batatia, Ilyes and Horton, Joshua T. and Pu, Yixuan and Kapil, Venkat and Witt, William C. and Magd{\u{a}}u, Ioan-Bogdan and Cole, Daniel J. and Cs{\'a}nyi, G{\'a}bor},
  title     = {{MACE-OFF}: Short-Range Transferable Machine Learning Force Fields for Organic Molecules},
  journal   = {J. Am. Chem. Soc.},
  year      = {2025},
  volume    = {147},
  pages     = {17598--17611},
  doi       = {10.1021/jacs.4c07099},
}

@article{Gong2025_BAMBOO_NMI,
  author    = {Gong, Sheng and Zhang, Yumin and Mu, Zhenliang and Pu, Zhichen and Wang, Hongyi and Han, Xu and Yu, Zhiao and Chen, Mengyi and Zheng, Tianze and Wang, Zhi and others},
  title     = {A Predictive Machine Learning Force-Field Framework for Liquid Electrolyte Development},
  journal   = {Nat. Mach. Intell.},
  year      = {2025},
  volume    = {7},
  pages     = {543--552},
  doi       = {10.1038/s42256-025-01009-7},
}

@article{Fu2023_arXiv,
  author    = {Fu, Xiang and Wu, Zhenghao and Wang, Wujie and Xie, Tian and Keten, Sinan and Gomez-Bombarelli, Rafael and Jaakkola, Tommi},
  title     = {Forces Are Not Enough: Benchmark and Critical Evaluation for Machine Learning Force Fields with Molecular Simulations},
  journal   = {arXiv preprint},
  year      = {2023},
  eprint    = {2210.07237},
  archiveprefix = {arXiv},
}

@article{Batatia2026_MACEPOLAR1,
  author    = {Batatia, Ilyes and Baldwin, William J. and Kuryla, Domantas and Hart, Joseph and Kasoar, Elliott and Elena, Alin M. and Moore, Harry and Gawkowski, Miko{\l}aj J. and Shi, Benjamin X. and Kapil, Venkat and Kourtis, Panagiotis and Magd{\u{a}}u, Ioan-Bogdan and Cs{\'a}nyi, G{\'a}bor},
  title     = {{MACE-POLAR-1}: A Polarisable Electrostatic Foundation Model for Molecular Chemistry},
  journal   = {arXiv preprint},
  year      = {2026},
  eprint    = {2602.19411},
  archiveprefix = {arXiv},
}

@article{Zhu2023_MBpol2023,
  author    = {Zhu, Xuanyu and Riera, Marc and Bull-Vulpe, Ethan F. and Paesani, Francesco},
  title     = {{MB-pol(2023)}: Sub-Chemical Accuracy for Water Simulations from the Gas to the Liquid Phase},
  journal   = {J. Chem. Theory Comput.},
  year      = {2023},
  volume    = {19},
  pages     = {3551--3566},
  doi       = {10.1021/acs.jctc.3c00326},
}

@article{Beckmann2025_JCIM_NNP_condensed,
  title={Transferable neural network potentials and condensed phase properties},
  author={Picha, Anna Katharina and Wieder, Marcus and Boresch, Stefan},
  journal={Journal of Chemical Information and Modeling},
  volume={65},
  number={18},
  pages={9483--9496},
  year={2025},
  publisher={ACS Publications}
}

@article{feng2025screening,
  title={Screening and design of aqueous zinc battery electrolytes based on the multimodal optimization of molecular simulation},
  author={Feng, Wei and Zhang, Luyan and Cheng, Yaobo and Wu, Jin and Wei, Chunguang and Zhang, Junwei and Yu, Kuang},
  journal={The Journal of Physical Chemistry Letters},
  volume={16},
  number={13},
  pages={3326--3335},
  year={2025},
  publisher={ACS Publications},
  url={https://pubs.acs.org/doi/10.1021/acs.jpclett.5c00341}
}

@article{Yue2021_JCP_shortrange,
  author    = {Yue, Shuwen and Muniz, Maria C. and Calegari Andrade, Marcos F. and Zhang, Linfeng and Car, Roberto and Panagiotopoulos, Athanassios Z.},
  title     = {When Do Short-Range Atomistic Machine-Learning Models Fall Short?},
  journal   = {J. Chem. Phys.},
  year      = {2021},
  volume    = {154},
  pages     = {034111},
  doi       = {10.1063/5.0031215},
}

@article{Anstine2023_JPCA_longrange,
  author    = {Anstine, Dylan M. and Isayev, Olexandr},
  title     = {Machine Learning Interatomic Potentials and Long-Range Physics},
  journal   = {J. Phys. Chem. A},
  year      = {2023},
  volume    = {127},
  pages     = {2417--2431},
  doi       = {10.1021/acs.jpca.2c06778},
}

@article{Schmidt2015_AccChemRes,
  author    = {Schmidt, J. R. and Yu, Kuang and McDaniel, Jesse G.},
  title     = {Transferable Next-Generation Force Fields from Simple Liquids to Complex Materials},
  journal   = {Acc. Chem. Res.},
  year      = {2015},
  volume    = {48},
  pages     = {548--556},
  doi       = {10.1021/ar500272n},
}

@article{Khaliullin2007_JPCA_ALMOEDA,
  author    = {Khaliullin, Rustam Z. and Cobar, Erika A. and Lochan, Rohini C. and Bell, Alexis T. and Head-Gordon, Martin},
  title     = {Unravelling the Origin of Intermolecular Interactions Using Absolutely Localized Molecular Orbitals},
  journal   = {J. Phys. Chem. A},
  year      = {2007},
  volume    = {111},
  pages     = {8753--8765},
  doi       = {10.1021/jp073685z},
}

@article{Misquitta2018_ISApol,
  author    = {Misquitta, Alston J. and Stone, Anthony J.},
  title     = {{ISA-Pol}: Distributed Polarizabilities and Dispersion Models from a Basis-Space Implementation of the Iterated Stockholder Atoms Procedure},
  journal   = {Theor. Chem. Acc.},
  year      = {2018},
  volume    = {137},
  pages     = {153},
  doi       = {10.1007/s00214-018-2371-4},
}

@article{Misquitta2014_JCTC_ISA,
  author    = {Misquitta, Alston J. and Stone, Anthony J. and Fazeli, Farhang},
  title     = {Distributed Multipoles from a Robust Basis-Space Implementation of the Iterated Stockholder Atoms Procedure},
  journal   = {J. Chem. Theory Comput.},
  year      = {2014},
  volume    = {10},
  pages     = {5405--5418},
  doi       = {10.1021/ct5008444},
}

@article{McDaniel2012_JPC_zeolite,
  author    = {McDaniel, Jesse G. and Yu, Kuang and Schmidt, J. R.},
  title     = {Ab Initio, Physically Motivated Force Fields for {CO$_2$} Adsorption in Zeolitic Imidazolate Frameworks},
  journal   = {J. Phys. Chem. C},
  year      = {2012},
  volume    = {116},
  pages     = {1892--1903},
  doi       = {10.1021/jp209335y},
}

@article{VanVleet2018_JCTC_anisotropy,
  author    = {Van Vleet, Mary J. and Misquitta, Alston J. and Schmidt, J. R.},
  title     = {New Angles on Standard Force Fields: Toward a General Approach for Treating Atomic-Level Anisotropy},
  journal   = {J. Chem. Theory Comput.},
  year      = {2018},
  volume    = {14},
  pages     = {739--758},
  doi       = {10.1021/acs.jctc.7b00851},
}

@article{Essmann1995_JCP,
  author    = {Essmann, Ulrich and Perera, Lalith and Berkowitz, Max L. and Darden, Tom and Lee, Hsing and Pedersen, Lee G.},
  title     = {A Smooth Particle Mesh {Ewald} Method},
  journal   = {J. Chem. Phys.},
  year      = {1995},
  volume    = {103},
  pages     = {8577--8593},
  doi       = {10.1063/1.470117},
}

@article{Huang2017_JCP_MPID,
  author    = {Huang, Jing and Simmonett, Andrew C. and Pickard, Frank C. and MacKerell, Alexander D. and Brooks, Bernard R.},
  title     = {Mapping the {Drude} Polarizable Force Field onto a Multipole and Induced Dipole Model},
  journal   = {J. Chem. Phys.},
  year      = {2017},
  volume    = {147},
  pages     = {161702},
  doi       = {10.1063/1.4984113},
}

@article{Wang2021_JPCL_sGNN,
  author    = {Wang, Xinyan and Xu, Yao and Zheng, Honghui and Yu, Kuang},
  title     = {A Scalable Graph Neural Network Method for Developing an Accurate Force Field of Large Flexible Organic Molecules},
  journal   = {J. Phys. Chem. Lett.},
  year      = {2021},
  volume    = {12},
  pages     = {7982--7987},
  doi       = {10.1021/acs.jpclett.1c02214},
}

@article{Zheng2025_ByteFF_DataDriven,
  author    = {Zheng, Tianze and Wang, Ailun and Han, Xu and Xia, Yu and Xu, Xingyuan and Zhan, Jiawei and Liu, Yu and Chen, Yang and Wang, Zhi and Wu, Xiaojie and others},
  title     = {Data-Driven Parametrization of Molecular Mechanics Force Fields for Expansive Chemical Space Coverage},
  journal   = {Chem. Sci.},
  year      = {2025},
  volume    = {16},
  pages     = {2730--2740},
  doi       = {10.1039/D4SC06640E},
}

@article{Cheng2024_JCTC_equivariant,
  author    = {Cheng, Zheng and Bi, Haoran and Liu, Shuai and Chen, Junmin and Misquitta, Alston J. and Yu, Kuang},
  title     = {Developing a Differentiable Long-Range Force Field for Proteins with {E(3)} Neural Network-Predicted Asymptotic Parameters},
  journal   = {J. Chem. Theory Comput.},
  year      = {2024},
  volume    = {20},
  pages     = {5598--5608},
  doi       = {10.1021/acs.jctc.4c00421},
}

@article{Wang2023_JCTC_DMFF,
  author    = {Wang, Xinyan and Li, Jichen and Yang, Lan and Chen, Feiyang and Wang, Yingze and Chang, Junhan and Chen, Junmin and Feng, Wei and Zhang, Linfeng and Yu, Kuang},
  title     = {{DMFF}: An Open-Source Automatic Differentiable Platform for Molecular Force Field Development and Molecular Dynamics Simulation},
  journal   = {J. Chem. Theory Comput.},
  year      = {2023},
  volume    = {19},
  pages     = {5897--5909},
  doi       = {10.1021/acs.jctc.2c01297},
}

@inproceedings{Schoenholz2020_NeurIPS_JAXMD,
  author    = {Schoenholz, Samuel S. and Cubuk, Ekin D.},
  title     = {{JAX-MD}: A Framework for Differentiable Physics},
  booktitle = {Advances in Neural Information Processing Systems},
  year      = {2020},
  volume    = {33},
  pages     = {11428--11441},
}

@article{Doerr2021_JCTC_TorchMD,
  author    = {Doerr, Stefan and Majewski, Maciej and P{\'e}rez, Adri{\`a} and Kr{\"a}mer, Andreas and Clementi, Cecilia and Noe, Frank and Giorgino, Toni and De Fabritiis, Gianni},
  title     = {{TorchMD}: A Deep Learning Framework for Molecular Simulations},
  journal   = {J. Chem. Theory Comput.},
  year      = {2021},
  volume    = {17},
  pages     = {2355--2363},
  doi       = {10.1021/acs.jctc.0c01343},
}

@article{Fuchs2024_arXiv_chemtrain,
  author    = {Fuchs, Paul and Thaler, Stephan and R{\"o}cken, Sebastian and Zavadlav, Julija},
  title     = {chemtrain: Learning Deep Potential Models via Automatic Differentiation and Statistical Physics},
  journal   = {arXiv preprint},
  year      = {2024},
  eprint    = {2408.15852},
  archiveprefix = {arXiv},
}

@article{Thaler2021_NatCommun,
  author    = {Thaler, Stephan and Zavadlav, Julija},
  title     = {Learning Neural Network Potentials from Experimental Data via Differentiable Trajectory Reweighting},
  journal   = {Nat. Commun.},
  year      = {2021},
  volume    = {12},
  pages     = {6884},
  doi       = {10.1038/s41467-021-27241-4},
}

@article{Han2025_NatCommun_IR,
  author    = {Han, Bin and Yu, Kuang},
  title     = {Refining Potential Energy Surface through Dynamical Properties via Differentiable Molecular Simulation},
  journal   = {Nat. Commun.},
  year      = {2025},
  volume    = {16},
  pages     = {816},
  doi       = {10.1038/s41467-025-56061-z},
}

@article{Jin2025_arXiv_phasediagram,
  author    = {Jin, Bin and Han, Bin and Feng, Wei and Yu, Kuang and Xu, Shenzhen},
  title     = {Automatic Refinement of Force Fields Based on Phase Diagrams},
  journal   = {arXiv preprint},
  year      = {2025},
  eprint    = {2510.16778},
  archiveprefix = {arXiv},
}

@article{Rocken2025_arXiv_DiffTTC,
  author    = {R{\"o}cken, Sebastian and Zavadlav, Julija and others},
  title     = {Refining Machine Learning Potentials through Thermodynamic Theory of Phase Transitions},
  journal   = {arXiv preprint},
  year      = {2025},
  eprint    = {2512.03974},
  archiveprefix = {arXiv},
}

@article{Metz2021_arXiv_gradients,
  author    = {Metz, Luke and Freeman, C. Daniel and Schoenholz, Samuel S. and Kachman, Tal},
  title     = {Gradients Are Not All You Need},
  journal   = {arXiv preprint},
  year      = {2021},
  eprint    = {2111.05803},
  archiveprefix = {arXiv},
}

@article{Yao2022_ChemRev,
  author    = {Yao, Nan and Chen, Xiang and Fu, Zhong-Heng and Zhang, Qiang},
  title     = {Applying Classical, Ab Initio, and Machine-Learning Molecular Dynamics Simulations to the Liquid Electrolyte for Rechargeable Batteries},
  journal   = {Chem. Rev.},
  year      = {2022},
  volume    = {122},
  pages     = {10970--11021},
  doi       = {10.1021/acs.chemrev.1c00904},
}

@article{Xu2004_ChemRev,
  author    = {Xu, Kang},
  title     = {Nonaqueous Liquid Electrolytes for Lithium-Based Rechargeable Batteries},
  journal   = {Chem. Rev.},
  year      = {2004},
  volume    = {104},
  pages     = {4303--4418},
  doi       = {10.1021/cr030203g},
}

@article{Meng2022_Science,
  author    = {Meng, Y. Shirley and Srinivasan, Venkat and Xu, Kang},
  title     = {Designing Better Electrolytes},
  journal   = {Science},
  year      = {2022},
  volume    = {378},
  pages     = {eabq3750},
  doi       = {10.1126/science.abq3750},
}

@article{Bedrov2019_ChemRev,
  author    = {Bedrov, Dmitry and Piquemal, Jean-Philip and Borodin, Oleg and MacKerell, Alexander D. and Roux, Beno{\^i}t and Schr{\"o}der, Christian},
  title     = {Molecular Dynamics Simulations of Ionic Liquids and Electrolytes Using Polarizable Force Fields},
  journal   = {Chem. Rev.},
  year      = {2019},
  volume    = {119},
  pages     = {7940--7995},
  doi       = {10.1021/acs.chemrev.8b00763},
}

@article{Dave2022_NatCommun_robot,
  author    = {Dave, Adarsh and Mitchell, Jared and Burke, Sven and Lin, Hai and Whitacre, Jay and Viswanathan, Venkatasubramanian},
  title     = {Autonomous Optimization of Non-Aqueous {Li}-Ion Battery Electrolytes via Robotic Experimentation and Machine Learning Coupling},
  journal   = {Nat. Commun.},
  year      = {2022},
  volume    = {13},
  pages     = {5454},
  doi       = {10.1038/s41467-022-32938-1},
}

@article{Kim2025_LES_Universal_JCTC,
author = {Kim, Dongjin and Wang, Xiaoyu and Vargas, Santiago and Zhong, Peichen and King, Daniel S and Inizan, Theo Jaffrelot and Cheng, Bingqing},
doi = {10.1021/acs.jctc.5c01400},
issn = {1549-9618},
journal = {Journal of Chemical Theory and Computation},
month = {dec},
number = {24},
pages = {12709--12724},
title = {{A Universal Augmentation Framework for Long-Range Electrostatics in Machine Learning Interatomic Potentials}},
url = {https://pubs.acs.org/doi/10.1021/acs.jctc.5c01400},
volume = {21},
year = {2025}
}

@software{jax2018github,
  author = {James Bradbury and Roy Frostig and Peter Hawkins and Matthew James Johnson and Chris Leary and Dougal Maclaurin and George Necula and Adam Paszke and Jake Vander{P}las and Skye Wanderman-{M}ilne and Qiao Zhang},
  title = {{JAX}: composable transformations of {P}ython+{N}um{P}y programs},
  url = {http://github.com/jax-ml/jax},
  version = {0.3.13},
  year = {2018},
}

@article{christiansen2025_DIMOS,
  title={Fast, modular, and differentiable framework for machine learning-enhanced molecular simulations},
  author={Christiansen, Henrik and Maruyama, Takashi and Errica, Federico and Zaverkin, Viktor and Takamoto, Makoto and Alesiani, Francesco},
  journal={The Journal of Chemical Physics},
  url = {https://pubs.aip.org/aip/jcp/article/163/18/182501/3371559},
  volume={163},
  number={18},
  year={2025},
  publisher={AIP Publishing}
}

@article{hannah2025searching,
  title={Searching for Ideal Electrolytes in the Molecular Universe},
  author={Hannah, Daniel and Zhang, Yumin and Li, Xinyu and Dong, Dengpan and Han, Joah and Park, Gyuleen and Gan, Hong and Liu, Bin and Liu, Kai and Hu, Qichao and others},
  journal={The Electrochemical Society Interface},
  volume={34},
  number={2},
  pages={35--38},
  year={2025},
  publisher={The Electrochemical Society},
  url={https://iopscience.iop.org/article/10.1149/2.f07252if}
}

@article{Ye2026_Interfacial,
	title = {Harnessing interfacial solvation structure for next-generation secondary batteries},
	volume = {11},
	issn = {2058-7546},
	url = {https://doi.org/10.1038/s41560-025-01937-z},
	doi = {10.1038/s41560-025-01937-z},
	number = {2},
	journal = {Nature Energy},
	author = {Ye, Chao and Tu, Shuibin and Zhang, Shao-Jian and Wang, Chunsheng and Qiao, Shi-Zhang},
	month = feb,
	year = {2026},
	pages = {167--175},
}

@article{Zhu2024_NatCommun_DiffMix,
  author    = {Zhu, Shang and others},
  title     = {Differentiable Modeling and Optimization of Non-Aqueous {Li}-Based Battery Electrolyte Solutions Using Geometric Deep Learning},
  journal   = {Nat. Commun.},
  year      = {2024},
  volume    = {15},
  pages     = {8649},
  doi       = {10.1038/s41467-024-51653-7},
}

@book{McQuarrie1976_StatMech,
  author    = {McQuarrie, Donald A.},
  title     = {Statistical Mechanics},
  publisher = {Harper \& Row},
  year      = {1976},
  address   = {New York},
}

@article{Zhang2020_EANN_spectra,
  author    = {Zhang, Yaolong and Ye, Sheng and Zhang, Jinxiao and Hu, Ce and Jiang, Jun and Jiang, Bin},
  title     = {Efficient and Accurate Simulations of Vibrational and Electronic Spectra with Symmetry-Preserving Neural Network Models for Tensorial Properties},
  journal   = {J. Phys. Chem. B},
  year      = {2020},
  volume    = {124},
  pages     = {7284--7290},
  doi       = {10.1021/acs.jpcb.0c06926},
}

@article{zhong_machine_2025,
    title = {Machine learning interatomic potential can infer electrical response},
    volume = {11},
    issn = {2057-3960},
    url = {https://www.nature.com/articles/s41524-025-01911-z},
    doi = {10.1038/s41524-025-01911-z},
    language = {english},
    number = {1},
    urldate = {2026-01-08},
    journal = {npj Computational Materials},
    author = {Zhong, Peichen and Kim, Dongjin and King, Daniel S. and Cheng, Bingqing},
    month = dec,
    year = {2025},
    pages = {384},
}

@article{musaelian_learning_2023,
    title = {Learning local equivariant representations for large-scale atomistic dynamics},
    volume = {14},
    issn = {2041-1723},
    url = {https://www.nature.com/articles/s41467-023-36329-y},
    doi = {10.1038/s41467-023-36329-y},
    language = {english},
    number = {1},
    urldate = {2026-03-18},
    journal = {Nature Communications},
    author = {Musaelian, Albert and Batzner, Simon and Johansson, Anders and Sun, Lixin and Owen, Cameron J. and Kornbluth, Mordechai and Kozinsky, Boris},
    month = feb,
    year = {2023},
    pages = {579},
}

@article{gan2025large,
  title={MatLLMSearch: Crystal Structure Discovery with Evolution-Guided Large Language Models},
  author={Gan, Jingru and Zhong, Peichen and Du, Yuanqi and Zhu, Yanqiao and Duan, Chenru and Wang, Haorui and Schwalbe-Koda, Daniel and Gomes, Carla P and Persson, Kristin A and Wang, Wei},
  journal={arXiv preprint arXiv:2502.20933},
  year={2025},
  url={https://arxiv.org/abs/2502.20933}
}

@article{du2025accelerating,
  title={{Accelerating Scientific Discovery with Autonomous Goal-evolving Agents}},
  author={Du, Yuanqi and Yu, Botao and Liu, Tianyu and Shen, Tony and Chen, Junwu and Rittig, Jan G and Sun, Kunyang and Zhang, Yikun and Song, Zhangde and Zhou, Bo and others},
  journal={arXiv preprint arXiv:2512.21782},
  year={2025},
  url={https://arxiv.org/abs/2512.21782}
}

@inproceedings{wangefficient2025,
  title={{Efficient Evolutionary Search Over Chemical Space with Large Language Models}},
  author={Wang, Haorui and Skreta, Marta and Ser, Cher Tian and Gao, Wenhao and Kong, Lingkai and Strieth-Kalthoff, Felix and Duan, Chenru and Zhuang, Yuchen and Yu, Yue and Zhu, Yanqiao and others},
  booktitle={The Thirteenth International Conference on Learning Representations},
  year={2025},
  url={https://arxiv.org/abs/2406.16976}
}

@inproceedings{liularge2024,
  title={{Large Language Models to Enhance Bayesian Optimization}},
  author={Liu, Tennison and Astorga, Nicol{\'a}s and Seedat, Nabeel and van der Schaar, Mihaela},
  booktitle={The Twelfth International Conference on Learning Representations},
  year={2024},
  url={https://arxiv.org/abs/2402.03921}
}

@inproceedings{kristiadi2024sober,
  title={{A Sober Look at LLMs for Material Discovery: Are They Actually Good for Bayesian Optimization Over Molecules?}},
  author={Kristiadi, Agustinus and Strieth-Kalthoff, Felix and Skreta, Marta and Poupart, Pascal and Aspuru-Guzik, Al{\'a}n and Pleiss, Geoff},
  booktitle={Proceedings of the 41st International Conference on Machine Learning},
  pages={25603--25622},
  year={2024},
  url={https://arxiv.org/abs/2402.05015}
}

@article{kim_high-entropy_2023,
    title = {High-entropy electrolytes for practical lithium metal batteries},
    volume = {8},
    issn = {2058-7546},
    url = {https://www.nature.com/articles/s41560-023-01280-1},
    doi = {10.1038/s41560-023-01280-1},
    number = {8},
    journal = {Nature Energy},
    publisher = {Springer US},
    author = {Kim, Sang Cheol and Wang, Jingyang and Xu, Rong and Zhang, Pu and Chen, Yuelang and Huang, Zhuojun and Yang, Yufei and Yu, Zhiao and Oyakhire, Solomon T. and Zhang, Wenbo and Greenburg, Louisa C. and Kim, Mun Sek and Boyle, David T. and Sayavong, Philaphon and Ye, Yusheng and Qin, Jian and Bao, Zhenan and Cui, Yi},
    month = jul,
    year = {2023},
    pages = {814--826},
}

@article{zhang_revolutionizing_2025,
    title = {Revolutionizing {Chemistry} and {Material} {Innovation}: {An} {Iterative} {Theoretical}-{Experimental} {Paradigm} {Leveraged} by {Robotic} {AI} {Chemists}},
    volume = {7},
    issn = {2096-5745},
    shorttitle = {Revolutionizing {Chemistry} and {Material} {Innovation}},
    url = {http://www.chinesechemsoc.org/doi/10.31635/ccschem.024.202404860},
    doi = {10.31635/ccschem.024.202404860},
    language = {english},
    number = {2},
    urldate = {2026-04-03},
    journal = {CCS Chemistry},
    author = {Zhang, Baicheng and Zhu, Zhuoying and Li, Huirong and Cao, Jiaqi and Jiang, Jun},
    month = feb,
    year = {2025},
    pages = {345--360},
}

@article{shen_unlocking_2026,
    title = {Unlocking azobenzene isomerization mechanisms \textit{via} an {LLM} agent-driven workflow integrating simulation, experiment, and machine learning},
    issn = {2041-6520, 2041-6539},
    url = {https://xlink.rsc.org/?DOI=D5SC08794E},
    doi = {10.1039/D5SC08794E},
    language = {english},
    urldate = {2026-04-03},
    journal = {Chemical Science},
    author = {Shen, Yixi and Wang, Ledu and Huang, Yan and Zhang, Xiaolong and Huang, Meng and Li, Huirong and He, Jing and Cai, Aoran and Wang, Yang and Smith, Pieter E. S. and Jiang, Jun and Zhu, Zhuoying and Chen, Linjiang},
    year = {2026},
    pages = {10.1039.D5SC08794E},
}

@article{sun_moses_2026,
    title = {{MOSES}: combining automated ontology construction with a multi-agent system for explainable chemical knowledge reasoning},
    volume = {2},
    issn = {3050-287X},
    shorttitle = {{MOSES}},
    url = {https://iopscience.iop.org/article/10.1088/3050-287X/ae3127},
    doi = {10.1088/3050-287X/ae3127},
    language = {english},
    number = {1},
    urldate = {2026-04-03},
    journal = {AI for Science},
    author = {Sun, Yingkai and Xu, Feiyang and Liang, Huadong and Fan, Xianghui and Wan, Guozhu and Zhong, Wenwan and Jiang, Jun and Li, Xin and Chen, Linjiang},
    month = mar,
    year = {2026},
    pages = {015001},
}

@article{song_multiagent-driven_2025,
    title = {A {Multiagent}-{Driven} {Robotic} {AI} {Chemist} {Enabling} {Autonomous} {Chemical} {Research} {On} {Demand}},
    volume = {147},
    copyright = {https://doi.org/10.15223/policy-029},
    issn = {0002-7863, 1520-5126},
    url = {https://pubs.acs.org/doi/10.1021/jacs.4c17738},
    doi = {10.1021/jacs.4c17738},
    language = {english},
    number = {15},
    urldate = {2026-04-03},
    journal = {Journal of the American Chemical Society},
    author = {Song, Tao and Luo, Man and Zhang, Xiaolong and Chen, Linjiang and Huang, Yan and Cao, Jiaqi and Zhu, Qing and Liu, Daobin and Zhang, Baicheng and Zou, Gang and Zhang, Guoqing and Zhang, Fei and Shang, Weiwei and Fu, Yao and Jiang, Jun and Luo, Yi},
    month = apr,
    year = {2025},
    pages = {12534--12545},
}

@article{campbell_mdcrow_2026,
    title = {{MDCrow}: automating molecular dynamics workflows with large language models},
    volume = {7},
    issn = {2632-2153},
    shorttitle = {{MDCrow}},
    url = {https://iopscience.iop.org/article/10.1088/2632-2153/ae4b07},
    doi = {10.1088/2632-2153/ae4b07},
    number = {2},
    urldate = {2026-04-03},
    journal = {Machine Learning: Science and Technology},
    author = {Campbell, Quintina and Cox, Sam and Medina, Jorge and Watterson, Brittany and White, Andrew D},
    month = apr,
    year = {2026},
    pages = {025037},
}

@misc{guilbert_dynamate_2025,
    title = {{DynaMate}: {An} {Autonomous} {Agent} for {Protein}-{Ligand} {Molecular} {Dynamics} {Simulations}},
    copyright = {Creative Commons Attribution 4.0 International},
    shorttitle = {{DynaMate}},
    url = {https://arxiv.org/abs/2512.10034},
    doi = {10.48550/ARXIV.2512.10034},
    urldate = {2026-04-03},
    publisher = {arXiv},
    author = {Guilbert, Salomé and Masschelein, Cassandra and Goumaz, Jeremy and Naida, Bohdan and Schwaller, Philippe},
    year = {2025},
    note = {Version Number: 1},
}

@article{ding_topolyagent_2026,
    title = {{ToPolyAgent}: {AI} agents for coarse-grained bead-spring topological polymer simulations},
    volume = {5},
    issn = {2635-098X},
    shorttitle = {{ToPolyAgent}},
    url = {https://xlink.rsc.org/?DOI=D5DD00471C},
    doi = {10.1039/D5DD00471C},
    number = {2},
    urldate = {2026-04-03},
    journal = {Digital Discovery},
    author = {Ding, Lijie and Carrillo, Jan-Michael and Do, Changwoo},
    year = {2026},
    pages = {901--909},
}

@misc{shi_mdagent2_2026,
    title = {{MDAgent2}: {Large} {Language} {Model} for {Code} {Generation} and {Knowledge} {Q}\&amp;{A} in {Molecular} {Dynamics}},
    copyright = {arXiv.org perpetual, non-exclusive license},
    shorttitle = {{MDAgent2}},
    url = {https://arxiv.org/abs/2601.02075},
    doi = {10.48550/ARXIV.2601.02075},
    urldate = {2026-04-03},
    publisher = {arXiv},
    author = {Shi, Zhuofan and A, Hubao and Shao, Yufei and Huang, Dongliang and An, Hongxu and Xin, Chunxiao and Shen, Haiyang and Wang, Zhenyu and Na, Yunshan and Huang, Gang and Jing, Xiang},
    year = {2026},
    note = {Version Number: 4},
}

@article{goodfellow_graph-based_2026,
    title = {Graph-{Based} {Internal} {Coordinate} {Analysis} for {Transition} {State} {Characterization}},
    volume = {22},
    copyright = {https://creativecommons.org/licenses/by/4.0/},
    issn = {1549-9618, 1549-9626},
    url = {https://pubs.acs.org/doi/10.1021/acs.jctc.5c02073},
    doi = {10.1021/acs.jctc.5c02073},
    language = {english},
    number = {5},
    urldate = {2026-04-03},
    journal = {Journal of Chemical Theory and Computation},
    author = {Goodfellow, Alister S. and Nguyen, Bao N.},
    month = mar,
    year = {2026},
    pages = {2348--2357},
}

@article{niblett_transferability_2025,
    title = {Transferability of {Data} {Sets} between {Machine}-{Learned} {Interatomic} {Potential} {Algorithms}},
    volume = {21},
    copyright = {https://creativecommons.org/licenses/by/4.0/},
    issn = {1549-9618, 1549-9626},
    url = {https://pubs.acs.org/doi/10.1021/acs.jctc.5c00272},
    doi = {10.1021/acs.jctc.5c00272},
    language = {english},
    number = {12},
    urldate = {2026-04-03},
    journal = {Journal of Chemical Theory and Computation},
    author = {Niblett, Samuel P. and Kourtis, Panagiotis and Magdău, Ioan-Bogdan and Grey, Clare P. and Csányi, Gábor},
    month = jun,
    year = {2025},
    pages = {6096--6112},
}

@article{Greener2025_PNAS_reversible,
  author    = {Greener, Joe G.},
  title     = {Reversibly differentiable simulation enables hybrid mechanistic and machine-learned approaches},
  journal   = {Proc. Natl. Acad. Sci. U.S.A.},
  year      = {2025},
  volume    = {122},
  number    = {38},
  pages     = {e2426058122},
  doi       = {10.1073/pnas.2426058122},
}

@article{Chen2026_PrecChem_redox,
author = {Chen, Yicheng and Cheng, Lixue and Jing, Yan and Zhong, Peichen},
doi = {10.1021/prechem.5c00258},
issn = {2771-9316},
journal = {Precision Chemistry},
month = {may},
number = {5},
pages = {612--621},
title = {{Benchmarking Foundation Potentials against Quantum Chemistry Methods for Predicting Molecular Redox Potentials}},
url = {https://pubs.acs.org/doi/10.1021/prechem.5c00258},
volume = {4},
year = {2026}
}

\end{document}